\documentclass[a4paper,10pt,twocolumn]{article}
\usepackage{inputenc}
\usepackage{tgtermes}
\usepackage{titlesec}
\usepackage{float}

\usepackage{soul}
\usepackage{amsmath,amssymb}
\usepackage{mathptmx}
\usepackage{bm,upgreek}
\usepackage{nicefrac}
\usepackage{graphicx}
\usepackage{xspace}
\usepackage[font=small]{caption}

\usepackage[super, comma, sort&compress]{natbib}

\makeatletter 
\renewcommand\@biblabel[1]{#1} 
\makeatother

\usepackage{upgreek} % for \upmu micro
\usepackage{braket}
\usepackage{authblk}
\usepackage{xspace}

\newcommand{\microns}{\xspace\ensuremath{\upmu\text{m}}\xspace}
\newcommand{\micron}{\ensuremath{\upmu\text{m}}\xspace}
\newcommand{\nm}{\xspace\ensuremath{\text{nm}}\xspace}

\newcommand{\tpa}{\xspace\ensuremath{\beta_\mathrm{TPA}}\xspace}
\newcommand{\tpawg}{\xspace\ensuremath{\alpha_\mathrm{TPA}}\xspace}

\usepackage[dvipsnames]{xcolor}
\newcommand{\new}[1]{#1}

\title{Mid-infrared quantum optics in silicon}
\date{}

\author[1,2]{Lawrence M. Rosenfeld}
\author[1,2]{Dominic A. Sulway}
\author[1]{Gary F. Sinclair}
\author[3]{Vikas Anant}
\author[1]{Mark~G.~Thompson}
\author[1]{John G. Rarity}
\author[1,*]{Joshua W. Silverstone}

\affil[1]{\small{Quantum Engineering Technology Labs, H. H. Wills Physics Laboratory and Department of Electrical and Electronic Engineering, University of Bristol, BS8 1FD, UK}}
\affil[2]{\small{Quantum Engineering Centre for Doctoral Training, H. H. Wills Physics Laboratory and Department of Electrical and Electronic Engineering, University of Bristol, BS8 1FD, UK}}
\affil[3]{\small{Photon Spot Inc., 142 West Olive Avenue, Monrovia, CA 91016, USA}}
\affil[*]{\small{josh.silverstone@bristol.ac.uk}}

\begin{document}
\definecolor{color0}{RGB}{0,0,0} % Base
\definecolor{color1}{RGB}{59,90,198} % author email, doi
\definecolor{color2}{RGB}{16,131,16} % Header, authors, table lines

% \maketitle
\twocolumn[
  \begin{@twocolumnfalse}
    \maketitle
    \vspace{-1.0cm}
\begin{abstract}
\noindent
Applied quantum optics stands to revolutionise many aspects of information technology, provided performance can be maintained when scaled up.
Silicon quantum photonics satisfies the scaling requirements of miniaturisation and manufacturability, but at 1.55\microns it suffers from unacceptable linear and nonlinear loss. %\cite{silverstone2016silicon}.
Here we show that, by translating silicon quantum photonics to the mid-infrared, a new quantum optics platform is created which can simultaneously maximise manufacturability and miniaturisation, while minimising loss.
We demonstrate the necessary platform components: photon-pair generation, single-photon detection, and high-visibility quantum interference, all at wavelengths beyond 2\microns.
Across various regimes, we observe a maximum net coincidence rate of $448~\pm~12$~Hz, a coincidence-to-accidental ratio of $25.7~\pm~1.1$, and, a net two-photon quantum interference visibility of $0.993~\pm~0.017$.
Mid-infrared silicon quantum photonics will bring new quantum applications within reach.%
\end{abstract}
\vspace{+0.5cm}
\end{@twocolumnfalse}
]

\vspace{1em}\noindent
The mid-infrared (MIR) is the energy band of vibrations. The molecular `fingerprint' region, 2--20\microns, is characterised by sharp molecular transitions, which lab-on-chip sensors can use to spectrally target molecular species in liquid and gaseous analytes\cite{sieger2016toward}. Lidar systems can exploit atmospheric transparency and reduced scintillation in the MIR\cite{corrigan2009quantum}, as well as the high-power handling and reduced phase error of MIR optical phased arrays for improved reliability\cite{prost2019mwir}. Much work has been done to bring integrated optics to the MIR, and silicon-on-insulator photonics dominates in the short-wavelength part of the band (the short-wave infrared), up to about 4\microns\cite{milovsevic2012silicon, nedeljkovic2013silicon, miller2017low, hattasan2012high}.  Here, silicon benefits from reduced two-photon absorption, facilitating nonlinear optical applications\cite{jalali2010silicon}: optical parametric oscillators and amplifiers\cite{liu2010mid}, supercontinuum sources\cite{kuyken2011mid,kou2018mid}, and frequency combs\cite{griffith2015silicon, yu2018silicon} have all been developed.

\begin{figure}[b!]
    \includegraphics[scale=1]{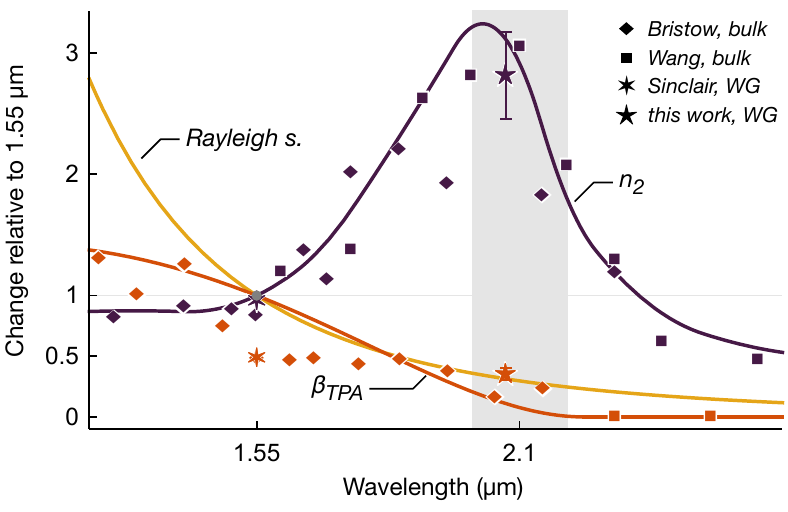}
    \caption{%
    Dispersion of key optical phenomena, relative to 1.55\microns values. Silicon intensity-dependent refractive index ($n_2$), and two-photon absorption coefficient ($\beta_\textrm{TPA}$) are shown, as well as simple Rayleigh scattering efficiency. Error bars represent one standard deviation of the mean of a Monte-Carlo-modelled distribution of system uncertainties. Values from Bristow \textit{et al.}\cite{bristow2007two} and Wang \textit{et al.}\cite{wang2013multi}, both measured using the bulk Z-scan technique, are plotted in diamonds and squares. Waveguided measurements from Sinclair \textit{et al.}\cite{sinclair2019temperature} and our work are plotted as  six- and five-pointed stars. Lines are: for $n_2$, a guide for the eye; for \tpa, a model for two-photon absorption\cite{garcia2006phonon}.%
       }
    \label{fig:motivation}
\end{figure}

Full quantum photonic technology has exquisite performance sensitivity, but stands to revolutionise how we measure, communicate, and ultimately process information\cite{o2009photonic}. It requires a huge scaling up for either integration or real-world deployment. As classical optics ventures into the MIR, quantum optics is close behind. Bulk-crystal photon-pair sources have been designed\cite{mccracken2018numerical}; experiments with one\cite{sua2017direct}, and two\cite{mancinelli2017mid} MIR photons have been shown, with detection provided by avalanche photodiodes and up-conversion.

\begin{figure*}[tb]
    \centering
    \includegraphics[scale = 1]{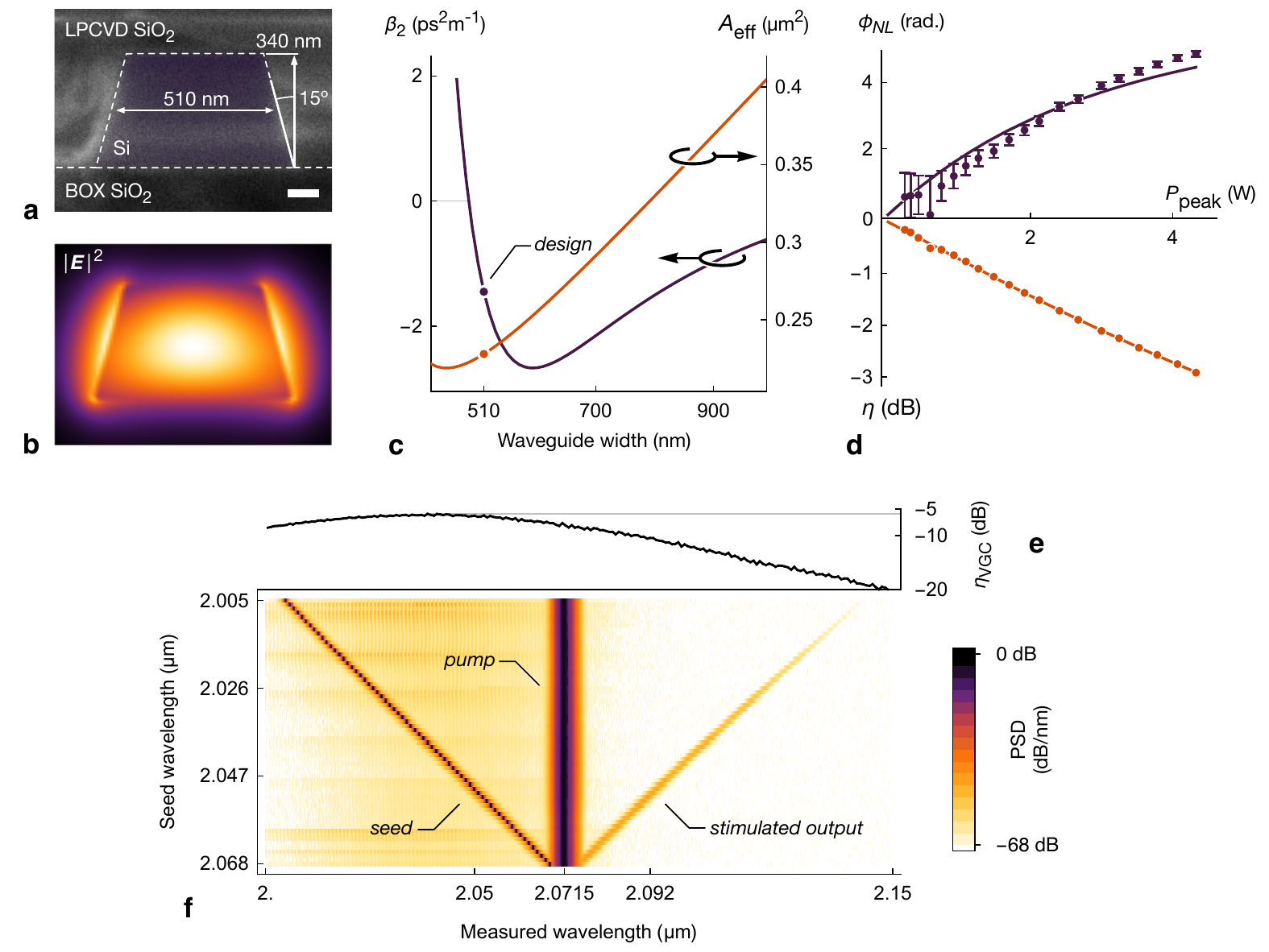}
    \caption{Waveguide design, simulation and experimental verification of phase-matching.  \textbf{a}, Scanning electron micrograph of the waveguide cross section. Scale bar $100~\text{nm}$. \textbf{b}, Simulation of the fundamental transverse electric mode electric field intensity at $\lambda = 2.071~\microns$. \textbf{c}, Simulations of the group velocity dispersion $\beta_2$ and effective modal area $A_\text{eff}$ varying the width of the source waveguide with a fixed height of $340~\text{nm}$ and side-wall angle of $15^{\circ}$. \textbf{d}, Nonlinear refraction and absorption. The maximum self-phase $\phi_{NL}$ and the normalised device transmission $\eta$ (decreasing due to residual two-photon absorption) are shown versus peak pump power. \textbf{e}, Single grating coupler transmission spectrum. \textbf{f}, Measured normalised power spectral density (PSD) of broadband stimulated four-wave mixing. A stimulating seed laser (continuous wave, tuneable, $\lambda\leq 2.071~\microns$) is swept on one side of the pulsed pump at $2.071~\microns$, while spectra are collected from an optical spectrum analyser, showing the stimulated output on the other ($\lambda\geq 2.071~\microns$).}
    \label{fig:prequantum}
\end{figure*}

Silicon photonics, operating mainly around the 1.5-\micron telecommunications band, has exploded in scale and functionality\cite{pavesi2016silicon}, and quantum silicon photonics has grown in tandem\cite{silverstone2016silicon}. In silicon, quantum-correlated photon pairs are scattered from a bright pump laser via the refractive nonlinearity, by spontaneous four-wave mixing\cite{sharping2006generation} (SFWM). Increasingly large interferometers have used these photons to power proof-of-concept quantum protocols\cite{harris2018linear, wang2018multidimensional, adcock2019programmable}, but to go beyond a handful of photons, ultra-low optical loss is needed. 

Propagation and device losses have steadily fallen\cite{cardenas2009low, lee2014low, benedikovic2015subwavelength, sheng2012compact, rouifed2016low}, but two-photon absorption (TPA) is intrinsic. TPA allows two photons to excite a crystal electron. It is stimulated absorption, growing with optical intensity. TPA clearly limits pump power, but it also limits the heralding efficiency of single photon sources\cite{husko2013multi}, and so is a fundamental limit to the large-scale viability of silicon quantum photonics\cite{silverstone2016silicon}. One common approach to avoiding TPA is to replace the guiding material (to e.g. silicon nitride\cite{ji2017ultra, ramelow2015silicon}). Beyond silicon's two-photon band edge, though, around 2.1\microns, two photons carry insufficient energy to excite a crystal electron, and TPA subsides. A resonant peak in the refractive nonlinearity, here, makes photon-pair sources more efficient, requiring less pump power. Conventional silica cladding remains transparent, and environmental black-body noise is not too large. Linear loss from Rayleigh scattering off etched waveguide side-walls\cite{hagan2017mechanisms, grillot2004size} is reduced at long wavelengths, and subwavelength features are more readily manufactured\cite{rouifed2016low, cheben2018subwavelength}. We plot the TPA coefficient \tpa, nonlinear refractive index $n_2$, and Rayleigh scattering cross-section, relative to their 1.55\microns values, in Fig.~\ref{fig:motivation}. In this article, we show that silicon photonics in the 2.1-\micron band is one route to ultra-low-loss quantum optics on a chip.

\begin{figure*}[ht!]
    \centering
    \includegraphics[scale = 1]{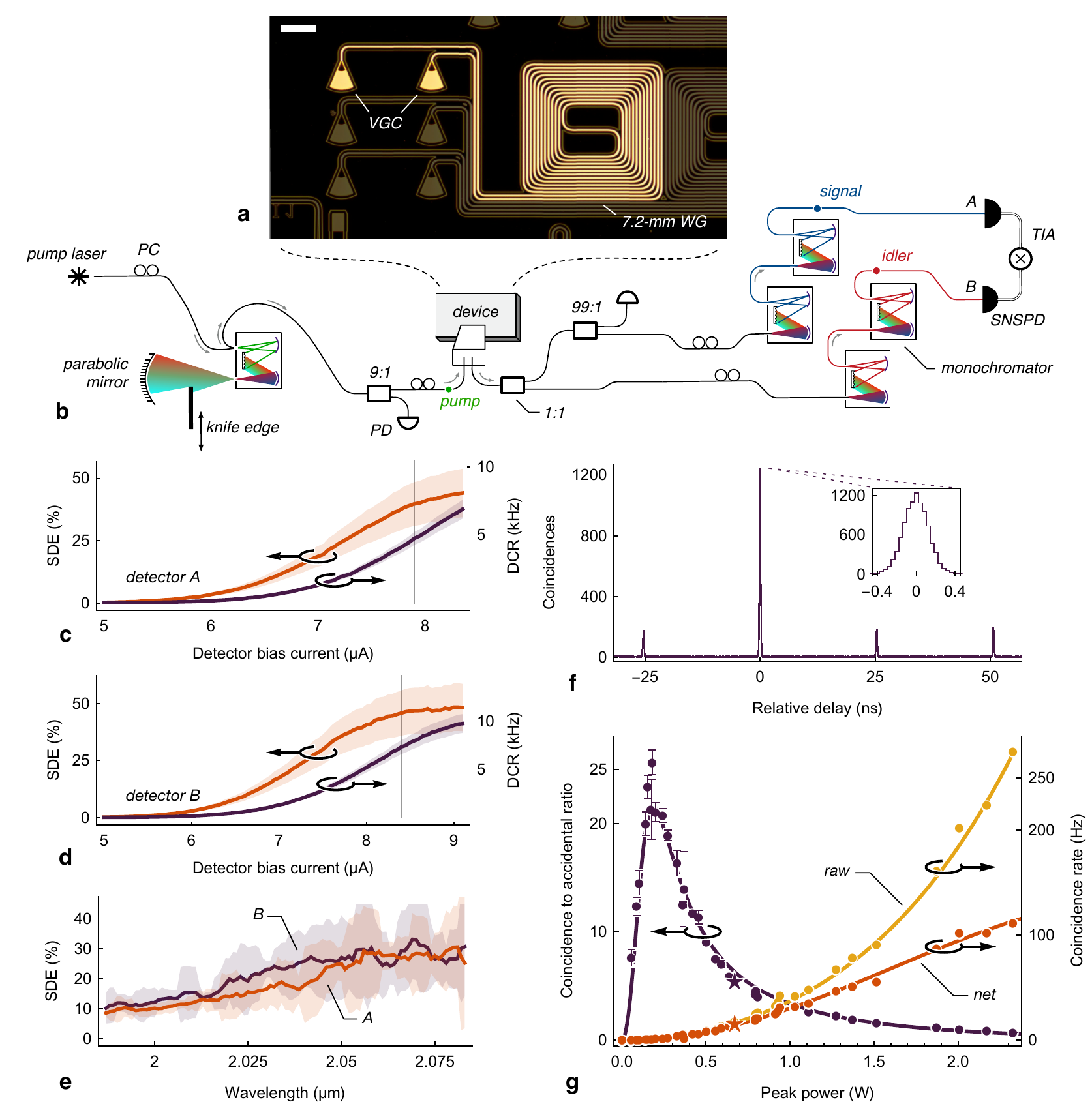}
    \caption{Measurement of correlated photons and characterisation of superconducting detectors:
    \textbf{a}, Dark field optical micrograph of the waveguide (WG) source with vertical grating couplers (VGC); scale bar $50~\microns$.
    \textbf{b}, Experimental configuration for correlated photon measurement. Polarisation controller (PC), input optical tap (9:1), photodiode (PD), beam splitter (1:1), output optical tap (99:1), superconducting nanowire single photon detector (SNSPD), time interval analyser (TIA).
    \textbf{c}, System detection efficiency (SDE) and dark count rate (DCR) with change in bias current measured at $\lambda = 2.07~\micron$ wavelength on detector $A$. Error bars are dominated by uncertainty in the number of launched photons.
    \textbf{d}, SDE and DCR for detector $B$. 
    \textbf{e}, Spectral response of detector efficiencies at a fixed bias current of $8.4$ and $7.9~\upmu\text{A}$ for detectors $A$ and $B$, respectively. A moving average window of five points has been applied to data and the error bars are the standard deviation of the points in the sampled moving average window.
    \textbf{f}, Sample coincidence histogram integrated for 540 seconds at $0.67$-W peak pump power. The peak at zero delay corresponds to photon pairs generated in the same spontaneous four-wave mixing event.
    \textbf{g}, Measured coincidence-to-accidental ratio (CAR), net and raw coincidence rate with varying launched pump power. Error bars are one standard deviation of the random error in the measurement. The sample histogram in part~\textbf{f} is indicated by a star.}
    \label{fig:photonpairs}
\end{figure*}

Any new quantum photonic platform needs: a source of quantum light, a way to detect that light, and quantum interference. We report on all three ingredients here. We design and characterise a silicon waveguide able to generate entangled photon pairs, centred on 2.07\microns, and use classical nonlinear optics to verify its design. We deploy a new detector system, optimised for the 2-\micron band, and verify its performance. We then drive SFWM in the designed waveguide and observe quantum-correlated photon pairs. Finally, we embed two such photon-pair sources in a reconfigurable on-chip interferometer and observe quantum interference.

\section*{\normalsize{Waveguide design for MIR SFWM}}
\label{sec:prequantum}

Spontaneous four-wave mixing, where two pump photons are scattered to higher and lower frequencies, conserves energy and momentum. For efficient SFWM, the  phase-matching condition must be satisfied. The total wave-vector mismatch, $\Delta k = \Delta k_\textrm{lin} - 2\gamma P,$ must be zero. Here, $P$ is the peak pump power in the waveguide, and $\gamma = \text k_{0}n_{2}/\text A_\text{eff}$ is the waveguide nonlinear parameter, with $n_2$ and $A_\text{eff}$ the nonlinear refractive index and effective modal area\cite{rukhlenko2012effective}, respectively. For frequencies near the pump, $\Delta k_\textrm{lin} = -\beta_{2}\,\Delta\omega^2,$ where $\beta_2 = \text{d}^2k/\text{d}\omega^2$ is the waveguide group-velocity dispersion (GVD), and $\Delta\omega$ is the pump-photon angular frequency detuning. For efficient four-wave mixing, $\beta_{2}\leq 0$, i.e. the GVD must be anomalous or zero.

We designed the waveguide shown in Fig.~\ref{fig:prequantum}a ($510\times340~\nm^2$ cross-section, $15^{\circ}$ side-wall angle) based on the variations of $\beta_2$ and $A_\mathrm{eff}$ with waveguide width shown in Fig.~\ref{fig:prequantum}c, at $\lambda=2.071~\microns$. The highly confined fundamental mode is shown in Fig.~\ref{fig:prequantum}b. To confirm the phase matching of our source, and estimate its bandwidth, we measure classical stimulated four-wave mixing\cite{liu2010mid, liscidini2013stimulated}. Figure~\ref{fig:prequantum}f shows a spectral map of the pump and stimulated emission, for various seed laser wavelengths. We observe phase-matched four-wave mixing over at least $60~\nm$, limited by grating coupler transmission, implying a very wide SFWM spectrum (Fig.~\ref{fig:prequantum}e).

All previous measurements of silicon's 2.1-\micron optical nonlinearity have been in bulk samples: here we confirm that the high nonlinear figure of merit measured in bulk also exists in nanoscale silicon waveguides. We use self-phase modulation and the Gerchberg-Saxton optical phase-retrieval method to estimate the waveguide nonlinearity\cite{sinclair2019temperature}; our results are summarised in Fig.~\ref{fig:prequantum}d with methods detailed in Supplementary Section~S2. We find an effective waveguide nonlinearity of $\gamma = 203\pm 26~\text{W}^{-1}\text{m}^{-1}$ ($n_2 = 15.3\pm 1.9\times 10^{-18}~\text{m}^2\,\text{W}^{-1}$ in bulk) and waveguide nonlinear absorption coefficient $\tpawg = 24.4\pm 3.3~\text{W}^{-1}\text{m}^{-1}$ ($\tpa=0.557\pm 0.07~\text{cm}\cdot\text{GW}^{-1}$ in bulk). Here, $\alpha_\text{TPA} = \partial\alpha/\partial P$ and $\beta_\text{TPA} = \partial\alpha/\partial I$, where $\alpha$ is the waveguide loss coefficient, and the intensity $I=P/A_\mathrm{eff}$ for a waveguided power $P$ with calculated $A_\mathrm{eff} = 0.228~\microns^2$. Our measurements on waveguides are in agreement with those in bulk\cite{wang2013multi, bristow2007two} (see Fig.~\ref{fig:motivation}). \new{
With the resonant $n_2$ enhancement, less pump is required for the same SFWM rate, and so less TPA ultimately results (see Supplementary Fig. S2). This is in addition to the modest reduction in \tpa which we observe.}

\section*{\normalsize{Observation of correlated photon pairs}}\label{sec:photons}

For a bright source of quantum-correlated photon pairs, we use SFWM. We start with a picosecond-pulsed pump laser, centred at $2.0715~\microns$, which we filter to a width of $1.0~\nm$ using a double-pass grating monochromator. Controlling polarisation, we inject this pump into the fundamental TE mode of the waveguide with a vertical grating coupler (VGC; $-7.3$ dB transmission). The waveguide, shown in Fig.~\ref{fig:photonpairs}a, is wrapped into a 7.2-mm square spiral (3.2 dB/cm propagation loss) with 10-\micron minimum radius Euler bends\cite{cherchi2013dramatic}.
Signal and idler photons are emitted in the same spatial mode, coupled off-chip, and separated probabilistically by a 1:1 fibre beam splitter. Both channels are tightly filtered with back-to-back free-space grating monochromators ($\sim$4.5 dB insertion loss per monochromator) to achieve the required $>100~\text{dB}$ pump rejection\cite{Piekarek:17, Savanier:16}. The signal and idler filters are $1.0$-nm wide, and separated from the pump by $\pm~1.46$~THz ($20.8$~nm). The experimental scheme is shown in Fig.~\ref{fig:photonpairs}b.

To detect single photons from SFWM, sensitive detectors are necessary. Superconducting nanowire single photon detectors (SNSPD) have unrivalled timing jitter, dark count rates (DCR), and system detection efficiency (SDE), and have shown sensitivity up to $5~\microns$\cite{marsili2012efficient}.  We incorporate superconducting nanowires into a dielectric stack optimised for absorption into the nanowire cavity at $\lambda=2.1~\microns$. The SNSPDs were fabricated from a 4-nm thick niobium nitride film, deposited using magnetron sputtering, and patterned  using electron-beam lithography into a meander of 100-nm-wide superconducting wire. The nanowires are fibre coupled with anti-reflection coated SM2000 fibre. A second SM2000-fibre span connects these detectors to our optical apparatus ($\sim30~\text{m}$; 1.49~dB and 1.63~dB loss per channel, for detectors $A$ and $B$ respectively). Operating at 780~mK, we find that the SDE of the two detectors plateaus with bias currents around $8~\upmu\text{A}$, with peak SDE values of $44\pm 10\%$ and $48\pm 10\%$, for detectors $A$ and $B$, respectively. In the following experiments, we operate with detector bias currents of $7.9~\upmu\text{A}$ and $8.1~\upmu\text{A}$. We measure a timing jitter of $216~\text{ps}$ (full width at half maximum) from photon cross-correlation. The dark count rate and efficiency as a function of bias at $\lambda=2.071~\microns$ is shown in Figs.~\ref{fig:photonpairs}c and \ref{fig:photonpairs}d. We use an attenuated tuneable laser to measure the SDE as a function of wavelength, and plot the results in Fig.~\ref{fig:photonpairs}e.

Photon pairs are detected in coincidence, with detector output pulses time correlated using standard coincidence counting logic. We observe a characteristic peak at zero relative delay from photons produced in the same SFWM event. A representative histogram is shown in Fig.~\ref{fig:photonpairs}f. We varied the launched optical power, fit each time correlation histogram with a Gaussian function, and integrated over 3 standard deviations (389~ps). These data are plotted in Fig.~\ref{fig:photonpairs}g. At low pump powers, we estimate an on-chip per-pulse generation efficiency of $11~\mathrm{MHz}/\mathrm{W}^2$, for watts of peak power, which at the repetition rate of our laser gives a generation probability of  $0.28/\mathrm{W}^2$ (peak). These estimated on-chip rates naturally exclude loss (see Methods). We measure a maximum net coincidence rate of $112\pm 3~\text{Hz}$, with peak coincidence to accidental ratio (CAR) of $25.7\pm1.1$, at a coincidence rate of $1.1~\text{Hz}$. We define $\mathrm{CAR}=(X_\mathrm{raw}-X_\mathrm{acc})/X_\mathrm{acc}$, where $X_\mathrm{raw}$ is the integrated coincidence count in the histogram central peak, and $X_\mathrm{acc}$ is the average integrated coincidence count in the side (accidental) peaks (see Supplementary Section~S4 and Fig.~S4 for singles rates). Dark counts limit the maximum CAR $\propto\eta/\mathrm{DCR}$, so higher system transmission ($\eta$) and lower DCR will offer further improvements to the CAR. On a longer 17.5-mm waveguide, we measured a peak net coincidence rate of $224~\text{Hz}$ (Supplementary Fig.~S2c). Due to the simple separation of the signal and idler with a beam splitter, the actual pair-production rates are $4\times$ the measured net rates. Thus, we observe true rates of $448~\text{Hz}$ and $896~\text{Hz}$ for the two waveguides, respectively (see Supplementary Section~S4).

\begin{figure*}[ht!]
    \centering
    \includegraphics[scale = 1]{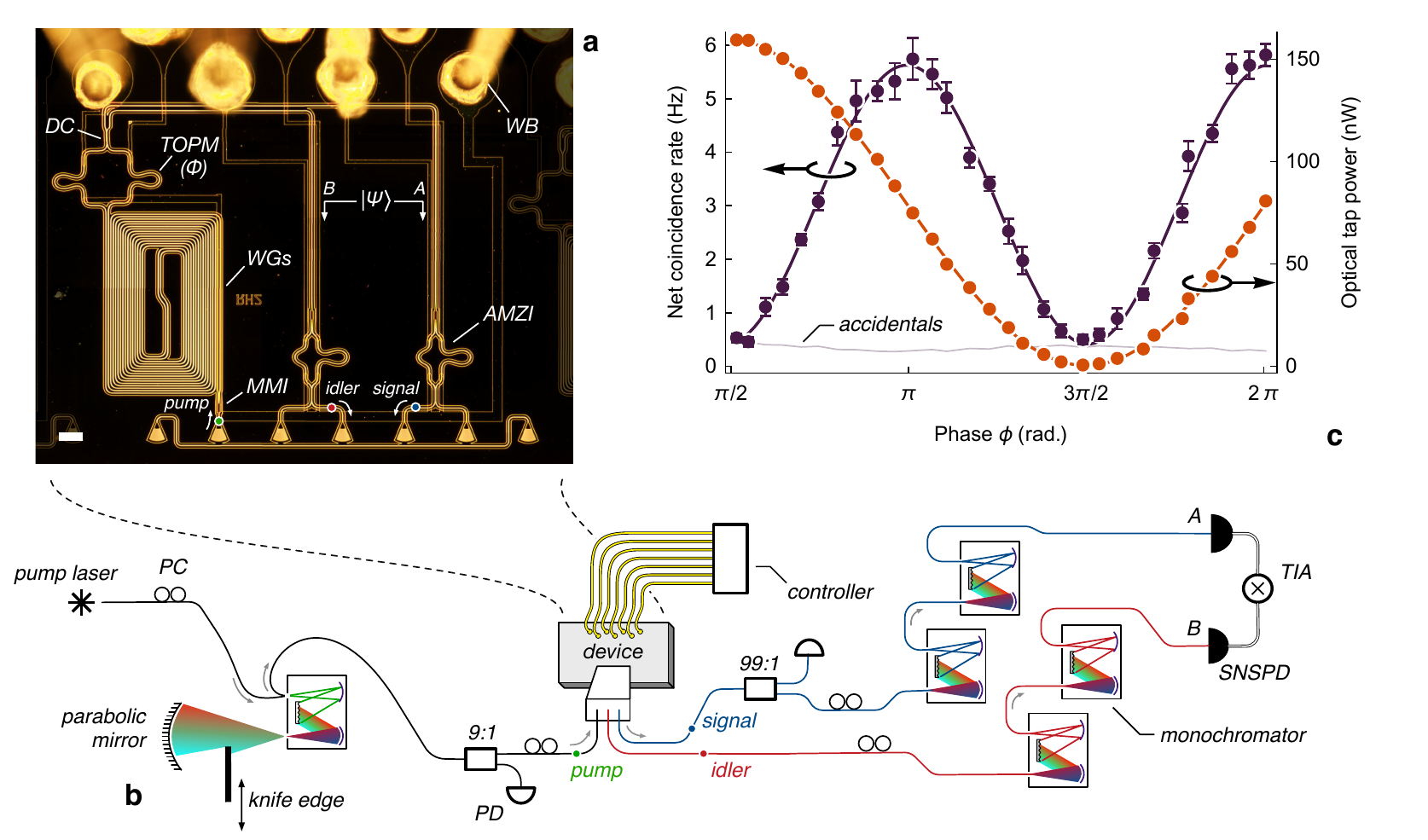}
    \caption{
    Experimental measurement of on-chip quantum interference:
    \textbf{a}, Dark-field optical micrograph of the time-reversed Hong-Ou-Mandel experiment. Multimode interference coupler (MMI), waveguides (7.4-mm, WGs), thermo-optic phase modulator (TOPM), directional coupler (DC), asymmetric Mach-Zehnder interferometer (AMZI), wirebond (WB). Scale bar 50 \micron.
    \textbf{b}, Experimental scheme. A pump laser is polarisation controlled (PC), filtered with a double-pass monochromator, and coupled into the waveguide circuit, with a monitor photodiode (PD) at the input tap (9:1). A controller provides DC-voltage control of the on-chip quantum state. The signal and idler photons are demultiplexed, filtered and then detected with superconducting nanowire detectors (SNSPD) and a time interval analyser (TIA).
    \textbf{c}, Quantum and classical interference fringes with varying on-chip phase $\phi$, with fitted accidental-subtracted (net) visibility $V = 0.993\pm0.017$.
    }
    \label{fig:quantuminterference}
\end{figure*}

\section*{\normalsize{On-chip quantum interference}}

The indistinguishability of generated photons, confirmed by high-visibility quantum interference between them, is an essential platform resource. The Hong-Ou-Mandel (HOM) effect causes two indistinguishable single photons to strictly bunch at the outputs of a balanced beam splitter\cite{hong1987measurement}. Here, we demonstrate on-chip quantum interference using a time-reversed HOM experiment\cite{silverstone2014chip}, with experimental setup and waveguide circuit shown in Figs. \ref{fig:quantuminterference}a and \ref{fig:quantuminterference}b.

After filtering the pump as before, we couple it onto the chip through a VGC. We set the on-chip peak pump power to $0.32~\text{W}$, corresponding to a CAR of $19.3$ and $\sim0.03$ pairs per pulse. The pump field is equally split between the two photon-pair sources, with a balanced $1\times2$ multimode interference (MMI) coupler. Both 7.39-mm sources are coherently pumped and the relative phase $\phi$ between the two arms is varied with a thermo-optic phase modulator (see Supplementary Section~S1 for details of integrated optics). The biphoton state then interferes on a balanced directional coupler ($R = 0.49 \pm 0.02$). Coherent pumping of both sources produces SFWM photon pairs in superposition, and the quantum state at the directional coupler output is
\begin{equation}
\label{eq:fringe}
\begin{split}
    \ket{\psi} \mkern5mu = \mkern5mu & \frac{\sin\phi}{\sqrt{2}}\big(\ket{0_s0_i}_A\ket{1_s1_i}_B - \ket{1_s1_i}_A\ket{0_s0_i}_B\big)\\ +\mkern5mu &\frac{\cos\phi}{\sqrt{2}}\big(\ket{1_s0_i}_A\ket{0_s1_i}_B + \ket{0_s1_i}_A\ket{1_s0_i}_B\big),
\end{split}
\end{equation}
with signal and idler frequency ($s,i$) and spatial modes ($A,B$) in subscript (see Supplementary Section~S5 for more details). SFWM photons are then frequency demultiplexed on-chip with asymmetric Mach-Zehnder interferometers. Off chip, we reject the pump laser, and use coincidence detection to estimate $|\bra{1_s0_i}_A\bra{0_s1_i}_B\ket{\psi}|^2$.%

We observe characteristic half-period interference fringes in the coincidences\cite{silverstone2014chip, matthews2009manipulation}, as the on-chip phase $\phi$ is varied, consistent with Eq.~\ref{eq:fringe} (see Fig.~\ref{fig:quantuminterference}c). This quantum interference has a net visibility of $V = 0.993 \pm 0.017$ (with $0.862\pm0.014$ raw). We calculate the visibility as $V = (X_\mathrm{max} - X_\mathrm{min})/(X_\mathrm{max} + X_\mathrm{min})$, where $X_\mathrm{max}$ and $X_\mathrm{min}$ are the maximum and minimum coincidence count rates of the sinusoidal fit. This compares favourably to performance at 1.55\microns on chip\cite{silverstone2014chip, he2015ultracompact, preble2015chip, jin2014chip}, and at 2.1\microns in bulk \cite{prabhakar2020two}. We observe coincidence rates of up to $5.5\pm 0.2~\text{Hz}$ at the interference peak. Simultaneously measuring all four chip outputs would double the observed rate. For perfect interference, the raw visibility is limited to $V \leq \mathrm{CAR}/(2 + \mathrm{CAR}) = 0.91$.

\section*{\normalsize{Discussion and conclusions}}

Despite operating in the 2-\micron band, we see from Fig.~\ref{fig:prequantum}d that two-photon absorption remains, albeit reduced from its strength at $1.55~\microns$. This is to be expected at room temperature and at $\lambda=2.07~\microns$, as silicon's indirect band gap gives a TPA cut-off around $2.21~\microns$ (Fig.~\ref{fig:motivation}). Our waveguided estimates of both nonlinear absorption and refraction are in broad agreement with literature values for bulk silicon\cite{wang2013multi, bristow2007two}. At low temperatures, a blue-shift in the band edge causes a blue-shift in the TPA cut-off\cite{cardona2004temperature}, to 2.15\microns. Future room-temperature experiments will benefit from 2.2-\micron laser development (e.g. semiconductor disk lasers\cite{kaspar2013recent}), while experiments integrating SNSPDs on-chip or in-package will benefit from this low-temperature shift of the TPA cut off.

Black-body radiation from our room-temperature apparatus was a source of noise in our measurements. As seen in Figs.~\ref{fig:photonpairs}c and \ref{fig:photonpairs}d, the `dark' count rate, collected with the lights and laser turned off, plateaus, rather than growing exponentially with bias, because environmental black-body photons nonetheless illuminate the detectors (see Supplementary Figs.~S3a,b). Black-body noise would be suppressed by a cold filter. Both CAR and raw fringe visibility would benefit from such a filter.

In demonstrating a bright source of photon pairs, efficient single-photon detectors, and high-visibility quantum interference, we have provided all the necessary ingredients for a dense, manufacturable, and high-performance platform for applied quantum optics. The mid-infrared presents a new approach to scalable optical quantum information processing in silicon.

\section*{\normalsize{Acknowledgements}}
We thank D. Bonneau, A. McMillan, B.D.J. Sayers, B. Kuyken, M. Nedeljkovic, K. Erotokritou, G. Taylor, and R.H. Hadfield for their valuable inputs throughout the long gestation of this work, and are grateful to L. Kling, and G.D. Marshall for their advice and support. The chip reported here was fabricated as part of the CORNERSTONE project, and we are especially grateful to C. Littlejohns for his efforts therein.

\section*{\normalsize{Disclosures}}
The authors declare no conflicts of interest.

\section*{\normalsize{Data Availability}}
Data and computer code that support the findings of this study are available at the University of Bristol's data repository, data.bris
%(Digital object identifier: XXXXXXXXXX)
. Other information is available from the authors upon reasonable request.

% Bibliography
\begin{footnotesize}
\bibliography{main}

\begin{thebibliography}{10}
\expandafter\ifx\csname url\endcsname\relax
  \def\url#1{\texttt{#1}}\fi
\expandafter\ifx\csname urlprefix\endcsname\relax\def\urlprefix{URL }\fi
\providecommand{\bibinfo}[2]{#2}
\providecommand{\eprint}[2][]{\url{#2}}

\bibitem{sieger2016toward}
\bibinfo{author}{Sieger, M.} \& \bibinfo{author}{Mizaikoff, B.}
\newblock \bibinfo{title}{Toward on-chip mid-infrared sensors.}
\newblock \emph{\bibinfo{journal}{Analytical Chemistry}}
  \textbf{\bibinfo{volume}{88}}, \bibinfo{pages}{5562} (\bibinfo{year}{2016}).

\bibitem{corrigan2009quantum}
\bibinfo{author}{Corrigan, P.}, \bibinfo{author}{Martini, R.},
  \bibinfo{author}{Whittaker, E.~A.} \& \bibinfo{author}{Bethea, C.}
\newblock \bibinfo{title}{Quantum cascade lasers and the kruse model in free
  space optical communication}.
\newblock \emph{\bibinfo{journal}{Optics Express}}
  \textbf{\bibinfo{volume}{17}}, \bibinfo{pages}{4355--4359}
  (\bibinfo{year}{2009}).

\bibitem{prost2019mwir}
\bibinfo{author}{Prost, M.} \emph{et~al.}
\newblock \bibinfo{title}{{MWIR} solid-state optical phased array beam steering
  using germanium-silicon photonic platform}.
\newblock In \emph{\bibinfo{booktitle}{Optical Fiber Communication
  Conference}}, \bibinfo{pages}{M4E--3} (\bibinfo{organization}{Optical Society
  of America}, \bibinfo{year}{2019}).

\bibitem{milovsevic2012silicon}
\bibinfo{author}{Milo{\v{s}}evi{\'c}, M.~M.} \emph{et~al.}
\newblock \bibinfo{title}{Silicon waveguides and devices for the mid-infrared}.
\newblock \emph{\bibinfo{journal}{Applied Physics Letters}}
  \textbf{\bibinfo{volume}{101}}, \bibinfo{pages}{121105}
  (\bibinfo{year}{2012}).

\bibitem{nedeljkovic2013silicon}
\bibinfo{author}{Nedeljkovic, M.} \emph{et~al.}
\newblock \bibinfo{title}{Silicon photonic devices and platforms for the
  mid-infrared}.
\newblock \emph{\bibinfo{journal}{Optical Materials Express}}
  \textbf{\bibinfo{volume}{3}}, \bibinfo{pages}{1205--1214}
  (\bibinfo{year}{2013}).

\bibitem{miller2017low}
\bibinfo{author}{Miller, S.~A.} \emph{et~al.}
\newblock \bibinfo{title}{Low-loss silicon platform for broadband mid-infrared
  photonics}.
\newblock \emph{\bibinfo{journal}{Optica}} \textbf{\bibinfo{volume}{4}},
  \bibinfo{pages}{707--712} (\bibinfo{year}{2017}).

\bibitem{hattasan2012high}
\bibinfo{author}{Hattasan, N.} \emph{et~al.}
\newblock \bibinfo{title}{High-efficiency {SOI} fiber-to-chip grating couplers
  and low-loss waveguides for the short-wave infrared}.
\newblock \emph{\bibinfo{journal}{IEEE Photonics Technology Letters}}
  \textbf{\bibinfo{volume}{24}}, \bibinfo{pages}{1536--1538}
  (\bibinfo{year}{2012}).

\bibitem{jalali2010silicon}
\bibinfo{author}{Jalali, B.}
\newblock \bibinfo{title}{Silicon photonics: Nonlinear optics in the
  mid-infrared}.
\newblock \emph{\bibinfo{journal}{Nature Photonics}}
  \textbf{\bibinfo{volume}{4}}, \bibinfo{pages}{506} (\bibinfo{year}{2010}).

\bibitem{liu2010mid}
\bibinfo{author}{Liu, X.}, \bibinfo{author}{Osgood~Jr, R.~M.},
  \bibinfo{author}{Vlasov, Y.~A.} \& \bibinfo{author}{Green, W.~M.}
\newblock \bibinfo{title}{Mid-infrared optical parametric amplifier using
  silicon nanophotonic waveguides}.
\newblock \emph{\bibinfo{journal}{Nature Photonics}}
  \textbf{\bibinfo{volume}{4}}, \bibinfo{pages}{557} (\bibinfo{year}{2010}).

\bibitem{kuyken2011mid}
\bibinfo{author}{Kuyken, B.} \emph{et~al.}
\newblock \bibinfo{title}{Mid-infrared to telecom-band supercontinuum
  generation in highly nonlinear silicon-on-insulator wire waveguides}.
\newblock \emph{\bibinfo{journal}{Optics Express}}
  \textbf{\bibinfo{volume}{19}}, \bibinfo{pages}{20172--20181}
  (\bibinfo{year}{2011}).

\bibitem{kou2018mid}
\bibinfo{author}{Kou, R.} \emph{et~al.}
\newblock \bibinfo{title}{Mid-{IR} broadband supercontinuum generation from a
  suspended silicon waveguide}.
\newblock \emph{\bibinfo{journal}{Optics Letters}}
  \textbf{\bibinfo{volume}{43}}, \bibinfo{pages}{1387--1390}
  (\bibinfo{year}{2018}).

\bibitem{griffith2015silicon}
\bibinfo{author}{Griffith, A.~G.} \emph{et~al.}
\newblock \bibinfo{title}{Silicon-chip mid-infrared frequency comb generation}.
\newblock \emph{\bibinfo{journal}{Nature Communications}}
  \textbf{\bibinfo{volume}{6}}, \bibinfo{pages}{6299} (\bibinfo{year}{2015}).

\bibitem{yu2018silicon}
\bibinfo{author}{Yu, M.} \emph{et~al.}
\newblock \bibinfo{title}{Silicon-chip-based mid-infrared dual-comb
  spectroscopy}.
\newblock \emph{\bibinfo{journal}{Nature Communications}}
  \textbf{\bibinfo{volume}{9}}, \bibinfo{pages}{1869} (\bibinfo{year}{2018}).

\bibitem{bristow2007two}
\bibinfo{author}{Bristow, A.~D.}, \bibinfo{author}{Rotenberg, N.} \&
  \bibinfo{author}{Van~Driel, H.~M.}
\newblock \bibinfo{title}{Two-photon absorption and kerr coefficients of
  silicon for 850--2200 nm}.
\newblock \emph{\bibinfo{journal}{Applied Physics Letters}}
  \textbf{\bibinfo{volume}{90}}, \bibinfo{pages}{191104}
  (\bibinfo{year}{2007}).

\bibitem{wang2013multi}
\bibinfo{author}{Wang, T.} \emph{et~al.}
\newblock \bibinfo{title}{Multi-photon absorption and third-order nonlinearity
  in silicon at mid-infrared wavelengths}.
\newblock \emph{\bibinfo{journal}{Optics Express}}
  \textbf{\bibinfo{volume}{21}}, \bibinfo{pages}{32192--32198}
  (\bibinfo{year}{2013}).

\bibitem{sinclair2019temperature}
\bibinfo{author}{Sinclair, G.~F.}, \bibinfo{author}{Tyler, N.~A.},
  \bibinfo{author}{Sahin, D.}, \bibinfo{author}{Barreto, J.} \&
  \bibinfo{author}{Thompson, M.~G.}
\newblock \bibinfo{title}{Temperature dependence of the kerr nonlinearity and
  two-photon absorption in a silicon waveguide at 1.55 $\upmu\text{m}$}.
\newblock \emph{\bibinfo{journal}{Physical Review Applied}}
  \textbf{\bibinfo{volume}{11}}, \bibinfo{pages}{044084}
  (\bibinfo{year}{2019}).

\bibitem{garcia2006phonon}
\bibinfo{author}{Garcia, H.} \& \bibinfo{author}{Kalyanaraman, R.}
\newblock \bibinfo{title}{Phonon-assisted two-photon absorption in the presence
  of a {DC}-field: the nonlinear {Franz--Keldysh} effect in indirect gap
  semiconductors}.
\newblock \emph{\bibinfo{journal}{Journal of Physics B: Atomic, Molecular and
  Optical Physics}} \textbf{\bibinfo{volume}{39}}, \bibinfo{pages}{2737}
  (\bibinfo{year}{2006}).

\bibitem{o2009photonic}
\bibinfo{author}{{O}'Brien, J.~L.}, \bibinfo{author}{Furusawa, A.} \&
  \bibinfo{author}{Vu{\v{c}}kovi{\'c}, J.}
\newblock \bibinfo{title}{Photonic quantum technologies}.
\newblock \emph{\bibinfo{journal}{Nature Photonics}}
  \textbf{\bibinfo{volume}{3}}, \bibinfo{pages}{687} (\bibinfo{year}{2009}).

\bibitem{mccracken2018numerical}
\bibinfo{author}{McCracken, R.~A.}, \bibinfo{author}{Graffitti, F.} \&
  \bibinfo{author}{Fedrizzi, A.}
\newblock \bibinfo{title}{Numerical investigation of mid-infrared single-photon
  generation}.
\newblock \emph{\bibinfo{journal}{JOSA B}} \textbf{\bibinfo{volume}{35}},
  \bibinfo{pages}{C38--C48} (\bibinfo{year}{2018}).

\bibitem{sua2017direct}
\bibinfo{author}{Sua, Y.~M.}, \bibinfo{author}{Fan, H.},
  \bibinfo{author}{Shahverdi, A.}, \bibinfo{author}{Chen, J.-Y.} \&
  \bibinfo{author}{Huang, Y.-P.}
\newblock \bibinfo{title}{Direct generation and detection of quantum correlated
  photons with 3.2 um wavelength spacing}.
\newblock \emph{\bibinfo{journal}{Scientific Reports}}
  \textbf{\bibinfo{volume}{7}}, \bibinfo{pages}{17494} (\bibinfo{year}{2017}).

\bibitem{mancinelli2017mid}
\bibinfo{author}{Mancinelli, M.} \emph{et~al.}
\newblock \bibinfo{title}{Mid-infrared coincidence measurements on twin photons
  at room temperature}.
\newblock \emph{\bibinfo{journal}{Nature Communications}}
  \textbf{\bibinfo{volume}{8}}, \bibinfo{pages}{15184} (\bibinfo{year}{2017}).

\bibitem{pavesi2016silicon}
\bibinfo{author}{Pavesi, L.} \& \bibinfo{author}{Lockwood, D.~J.}
\newblock \emph{\bibinfo{title}{Silicon photonics III: Systems and
  applications}}, vol. \bibinfo{volume}{122} (\bibinfo{publisher}{Springer
  Science \& Business Media}, \bibinfo{year}{2016}).

\bibitem{silverstone2016silicon}
\bibinfo{author}{Silverstone, J.~W.}, \bibinfo{author}{Bonneau, D.},
  \bibinfo{author}{O'Brien, J.~L.} \& \bibinfo{author}{Thompson, M.~G.}
\newblock \bibinfo{title}{Silicon quantum photonics}.
\newblock \emph{\bibinfo{journal}{IEEE Journal of Selected Topics in Quantum
  Electronics}} \textbf{\bibinfo{volume}{22}}, \bibinfo{pages}{390--402}
  (\bibinfo{year}{2016}).

\bibitem{sharping2006generation}
\bibinfo{author}{Sharping, J.~E.} \emph{et~al.}
\newblock \bibinfo{title}{{Generation of correlated photons in nanoscale
  silicon waveguides.}}
\newblock \emph{\bibinfo{journal}{Optics Express}}
  \textbf{\bibinfo{volume}{14}}, \bibinfo{pages}{12388--12393}
  (\bibinfo{year}{2006}).

\bibitem{harris2018linear}
\bibinfo{author}{Harris, N.~C.} \emph{et~al.}
\newblock \bibinfo{title}{{Linear programmable nanophotonic processors}}.
\newblock \emph{\bibinfo{journal}{Optica}} \textbf{\bibinfo{volume}{5}},
  \bibinfo{pages}{1623} (\bibinfo{year}{2018}).

\bibitem{wang2018multidimensional}
\bibinfo{author}{Wang, J.} \emph{et~al.}
\newblock \bibinfo{title}{Multidimensional quantum entanglement with
  large-scale integrated optics}.
\newblock \emph{\bibinfo{journal}{Science}} \textbf{\bibinfo{volume}{360}},
  \bibinfo{pages}{285--291} (\bibinfo{year}{2018}).

\bibitem{adcock2019programmable}
\bibinfo{author}{Adcock, J.~C.}, \bibinfo{author}{Vigliar, C.},
  \bibinfo{author}{Santagati, R.}, \bibinfo{author}{Silverstone, J.~W.} \&
  \bibinfo{author}{Thompson, M.~G.}
\newblock \bibinfo{title}{Programmable four-photon graph states on a silicon
  chip}.
\newblock \emph{\bibinfo{journal}{Nature Communications}}
  \textbf{\bibinfo{volume}{10}}, \bibinfo{pages}{3528} (\bibinfo{year}{2019}).

\bibitem{cardenas2009low}
\bibinfo{author}{Cardenas, J.} \emph{et~al.}
\newblock \bibinfo{title}{Low loss etchless silicon photonic waveguides}.
\newblock \emph{\bibinfo{journal}{Optics Express}}
  \textbf{\bibinfo{volume}{17}}, \bibinfo{pages}{4752--4757}
  (\bibinfo{year}{2009}).

\bibitem{lee2014low}
\bibinfo{author}{Lee, D.~H.} \emph{et~al.}
\newblock \bibinfo{title}{Low-loss silicon waveguides with sidewall roughness
  reduction using a {SiO2} hard mask and fluorine-based dry etching}.
\newblock \emph{\bibinfo{journal}{Journal of Micromechanics and
  Microengineering}} \textbf{\bibinfo{volume}{25}}, \bibinfo{pages}{015003}
  (\bibinfo{year}{2014}).

\bibitem{benedikovic2015subwavelength}
\bibinfo{author}{Benedikovic, D.} \emph{et~al.}
\newblock \bibinfo{title}{Subwavelength index engineered surface grating
  coupler with sub-decibel efficiency for 220-nm silicon-on-insulator
  waveguides}.
\newblock \emph{\bibinfo{journal}{Optics Express}}
  \textbf{\bibinfo{volume}{23}}, \bibinfo{pages}{22628--22635}
  (\bibinfo{year}{2015}).

\bibitem{sheng2012compact}
\bibinfo{author}{Sheng, Z.} \emph{et~al.}
\newblock \bibinfo{title}{A compact and low-loss {MMI} coupler fabricated with
  {CMOS} technology}.
\newblock \emph{\bibinfo{journal}{IEEE Photonics Journal}}
  \textbf{\bibinfo{volume}{4}}, \bibinfo{pages}{2272--2277}
  (\bibinfo{year}{2012}).

\bibitem{rouifed2016low}
\bibinfo{author}{Rouifed, M.-S.} \emph{et~al.}
\newblock \bibinfo{title}{Low loss {SOI} waveguides and {MMIs} at the {MIR}
  wavelength of $2~\upmu \text{m}$}.
\newblock \emph{\bibinfo{journal}{IEEE Photonics Technology Letters}}
  \textbf{\bibinfo{volume}{28}}, \bibinfo{pages}{2827--2829}
  (\bibinfo{year}{2016}).

\bibitem{husko2013multi}
\bibinfo{author}{Husko, C.~A.} \emph{et~al.}
\newblock \bibinfo{title}{Multi-photon absorption limits to heralded single
  photon sources}.
\newblock \emph{\bibinfo{journal}{Scientific Reports}}
  \textbf{\bibinfo{volume}{3}}, \bibinfo{pages}{3087} (\bibinfo{year}{2013}).

\bibitem{ji2017ultra}
\bibinfo{author}{Ji, X.} \emph{et~al.}
\newblock \bibinfo{title}{Ultra-low-loss on-chip resonators with sub-milliwatt
  parametric oscillation threshold}.
\newblock \emph{\bibinfo{journal}{Optica}} \textbf{\bibinfo{volume}{4}},
  \bibinfo{pages}{619--624} (\bibinfo{year}{2017}).

\bibitem{ramelow2015silicon}
\bibinfo{author}{Ramelow, S.} \emph{et~al.}
\newblock \bibinfo{title}{Silicon-nitride platform for narrowband entangled
  photon generation}.
\newblock \emph{\bibinfo{journal}{arXiv preprint arXiv:1508.04358}}
  (\bibinfo{year}{2015}).

\bibitem{hagan2017mechanisms}
\bibinfo{author}{Hagan, D.~E.} \& \bibinfo{author}{Knights, A.~P.}
\newblock \bibinfo{title}{Mechanisms for optical loss in {SOI} waveguides for
  mid-infrared wavelengths around 2 $\upmu\text{m}$}.
\newblock \emph{\bibinfo{journal}{Journal of Optics}}
  \textbf{\bibinfo{volume}{19}} (\bibinfo{year}{2017}).

\bibitem{grillot2004size}
\bibinfo{author}{Grillot, F.}, \bibinfo{author}{Vivien, L.},
  \bibinfo{author}{Laval, S.}, \bibinfo{author}{Pascal, D.} \&
  \bibinfo{author}{Cassan, E.}
\newblock \bibinfo{title}{Size influence on the propagation loss induced by
  sidewall roughness in ultrasmall soi waveguides}.
\newblock \emph{\bibinfo{journal}{IEEE Photonics Technology Letters}}
  \textbf{\bibinfo{volume}{16}}, \bibinfo{pages}{1661--1663}
  (\bibinfo{year}{2004}).

\bibitem{cheben2018subwavelength}
\bibinfo{author}{Cheben, P.}, \bibinfo{author}{Halir, R.},
  \bibinfo{author}{Schmid, J.~H.}, \bibinfo{author}{Atwater, H.~A.} \&
  \bibinfo{author}{Smith, D.~R.}
\newblock \bibinfo{title}{Subwavelength integrated photonics}.
\newblock \emph{\bibinfo{journal}{Nature}} \textbf{\bibinfo{volume}{560}},
  \bibinfo{pages}{565} (\bibinfo{year}{2018}).

\bibitem{rukhlenko2012effective}
\bibinfo{author}{Rukhlenko, I.~D.}, \bibinfo{author}{Premaratne, M.} \&
  \bibinfo{author}{Agrawal, G.~P.}
\newblock \bibinfo{title}{Effective mode area and its optimization in
  silicon-nanocrystal waveguides}.
\newblock \emph{\bibinfo{journal}{Optics letters}}
  \textbf{\bibinfo{volume}{37}}, \bibinfo{pages}{2295--2297}
  (\bibinfo{year}{2012}).

\bibitem{liscidini2013stimulated}
\bibinfo{author}{Liscidini, M.} \& \bibinfo{author}{Sipe, J.}
\newblock \bibinfo{title}{Stimulated emission tomography}.
\newblock \emph{\bibinfo{journal}{Physical Review Letters}}
  \textbf{\bibinfo{volume}{111}}, \bibinfo{pages}{193602}
  (\bibinfo{year}{2013}).

\bibitem{cherchi2013dramatic}
\bibinfo{author}{Cherchi, M.}, \bibinfo{author}{Ylinen, S.},
  \bibinfo{author}{Harjanne, M.}, \bibinfo{author}{Kapulainen, M.} \&
  \bibinfo{author}{Aalto, T.}
\newblock \bibinfo{title}{{Dramatic size reduction of waveguide bends on a
  micron-scale silicon photonic platform}}.
\newblock \emph{\bibinfo{journal}{Optics Express}}
  \textbf{\bibinfo{volume}{21}}, \bibinfo{pages}{17814--17823}
  (\bibinfo{year}{2013}).

\bibitem{Piekarek:17}
\bibinfo{author}{Piekarek, M.} \emph{et~al.}
\newblock \bibinfo{title}{High-extinction ratio integrated photonic filters for
  silicon quantum photonics}.
\newblock \emph{\bibinfo{journal}{Optics Letters}}
  \textbf{\bibinfo{volume}{42}}, \bibinfo{pages}{815--818}
  (\bibinfo{year}{2017}).

\bibitem{Savanier:16}
\bibinfo{author}{Savanier, M.}, \bibinfo{author}{Kumar, R.} \&
  \bibinfo{author}{Mookherjea, S.}
\newblock \bibinfo{title}{Photon pair generation from compact silicon microring
  resonators using microwatt-level pump powers}.
\newblock \emph{\bibinfo{journal}{Optics Express}}
  \textbf{\bibinfo{volume}{24}}, \bibinfo{pages}{3313--3328}
  (\bibinfo{year}{2016}).

\bibitem{marsili2012efficient}
\bibinfo{author}{Marsili, F.} \emph{et~al.}
\newblock \bibinfo{title}{Efficient single photon detection from 500 nm to 5
  $\upmu\text{m}$ wavelength}.
\newblock \emph{\bibinfo{journal}{Nano letters}} \textbf{\bibinfo{volume}{12}},
  \bibinfo{pages}{4799--4804} (\bibinfo{year}{2012}).

\bibitem{hong1987measurement}
\bibinfo{author}{Hong, C.-K.}, \bibinfo{author}{Ou, Z.-Y.} \&
  \bibinfo{author}{Mandel, L.}
\newblock \bibinfo{title}{Measurement of subpicosecond time intervals between
  two photons by interference}.
\newblock \emph{\bibinfo{journal}{Physical Review Letters}}
  \textbf{\bibinfo{volume}{59}}, \bibinfo{pages}{2044} (\bibinfo{year}{1987}).

\bibitem{silverstone2014chip}
\bibinfo{author}{Silverstone, J.~W.} \emph{et~al.}
\newblock \bibinfo{title}{On-chip quantum interference between silicon
  photon-pair sources}.
\newblock \emph{\bibinfo{journal}{Nature Photonics}}
  \textbf{\bibinfo{volume}{8}}, \bibinfo{pages}{104} (\bibinfo{year}{2014}).

\bibitem{matthews2009manipulation}
\bibinfo{author}{Matthews, J.~C.}, \bibinfo{author}{Politi, A.},
  \bibinfo{author}{Stefanov, A.} \& \bibinfo{author}{{O'Brien}, J.~L.}
\newblock \bibinfo{title}{Manipulation of multiphoton entanglement in waveguide
  quantum circuits}.
\newblock \emph{\bibinfo{journal}{Nature Photonics}}
  \textbf{\bibinfo{volume}{3}}, \bibinfo{pages}{346} (\bibinfo{year}{2009}).

\bibitem{he2015ultracompact}
\bibinfo{author}{He, J.} \emph{et~al.}
\newblock \bibinfo{title}{Ultracompact quantum splitter of degenerate photon
  pairs}.
\newblock \emph{\bibinfo{journal}{Optica}} \textbf{\bibinfo{volume}{2}},
  \bibinfo{pages}{779--782} (\bibinfo{year}{2015}).

\bibitem{preble2015chip}
\bibinfo{author}{Preble, S.~F.} \emph{et~al.}
\newblock \bibinfo{title}{On-chip quantum interference from a single silicon
  ring-resonator source}.
\newblock \emph{\bibinfo{journal}{Physical Review Applied}}
  \textbf{\bibinfo{volume}{4}}, \bibinfo{pages}{021001} (\bibinfo{year}{2015}).

\bibitem{jin2014chip}
\bibinfo{author}{Jin, H.} \emph{et~al.}
\newblock \bibinfo{title}{On-chip generation and manipulation of entangled
  photons based on reconfigurable lithium-niobate waveguide circuits}.
\newblock \emph{\bibinfo{journal}{Physical review letters}}
  \textbf{\bibinfo{volume}{113}}, \bibinfo{pages}{103601}
  (\bibinfo{year}{2014}).

\bibitem{prabhakar2020two}
\bibinfo{author}{Prabhakar, S.} \emph{et~al.}
\newblock \bibinfo{title}{Two-photon quantum interference and entanglement at
  2.1 $\upmu$m}.
\newblock \emph{\bibinfo{journal}{Science Advances}}
  \textbf{\bibinfo{volume}{6}} (\bibinfo{year}{2020}).

\bibitem{cardona2004temperature}
\bibinfo{author}{Cardona, M.}, \bibinfo{author}{Meyer, T.~A.} \&
  \bibinfo{author}{Thewalt, M. L.~W.}
\newblock \bibinfo{title}{{Temperature Dependence of the Energy Gap of
  Semiconductors in the Low-Temperature Limit}}.
\newblock \emph{\bibinfo{journal}{Physical Review Letters}}
  \textbf{\bibinfo{volume}{92}}, \bibinfo{pages}{623} (\bibinfo{year}{2004}).

\bibitem{kaspar2013recent}
\bibinfo{author}{Kaspar, S.} \emph{et~al.}
\newblock \bibinfo{title}{{Recent Advances in 2-$\mu$m GaSb-Based Semiconductor
  Disk Laser{\textemdash}Power Scaling, Narrow-Linewidth and Short-Pulse
  Operation}}.
\newblock \emph{\bibinfo{journal}{IEEE Journal of Selected Topics in Quantum
  Electronics}} \textbf{\bibinfo{volume}{19}},
  \bibinfo{pages}{1501908--1501908} (\bibinfo{year}{2013}).

\bibitem{mode2018lumerical}
\bibinfo{title}{{Lumerical Mode Solutions}} (\bibinfo{year}{2018}).

\bibitem{Fienup:82}
\bibinfo{author}{Fienup, J.~R.}
\newblock \bibinfo{title}{Phase retrieval algorithms: a comparison}.
\newblock \emph{\bibinfo{journal}{Applied Optics}}
  \textbf{\bibinfo{volume}{21}}, \bibinfo{pages}{2758--2769}
  (\bibinfo{year}{1982}).

\bibitem{PhysRevLett.30.901}
\bibinfo{author}{Reintjes, J.~F.} \& \bibinfo{author}{McGroddy, J.~C.}
\newblock \bibinfo{title}{Indirect two-photon transitions in si at 1.06
  $\upmu\text{m}$}.
\newblock \emph{\bibinfo{journal}{Physical Review Letters}}
  \textbf{\bibinfo{volume}{30}}, \bibinfo{pages}{901--903}
  (\bibinfo{year}{1973}).

\bibitem{rarity1987absolute}
\bibinfo{author}{Rarity, J.~G.}, \bibinfo{author}{Ridley, K.~D.} \&
  \bibinfo{author}{Tapster, P.~R.}
\newblock \bibinfo{title}{{Absolute measurement of detector quantum efficiency
  using parametric downconversion}}.
\newblock \emph{\bibinfo{journal}{Applied optics}}
  \textbf{\bibinfo{volume}{26}}, \bibinfo{pages}{4616--4619}
  (\bibinfo{year}{1987}).

\end{thebibliography}
\end{footnotesize}

\section*{\textbf{Methods}}
\begin{footnotesize}
\paragraph{\footnotesize{Dispersion calculation}}
Finite difference eigenmode simulations were performed to determine the optimal waveguide geometry \cite{ mode2018lumerical}. The waveguide height was fixed to $340~\text{nm}$, with a $15^{\circ}$ sidewall angle. The dispersion was calculated with a fractional frequency offset of 0.0001 to obtain the group velocity dispersion for the fundamental transverse electric mode at a centre wavelength of $\lambda = 2.071~\microns$. A waveguide with a cross-sectional area of $510\times340~\text{nm}^{2}$ simultaneously achieves anomalous group-velocity dispersion ($\beta_{2}\leq0$) required for phase matching, while minimising the effective modal area, see Fig.~\ref{fig:prequantum}c.
\paragraph{\footnotesize{Device fabrication}}
The device in this report was a $340~\text{nm}$ crystalline silicon(100)-oriented silicon-on-insulator platform, with a $2~\micron$ buried oxide and LPCVD $\text{SiO}_2$ $1~\micron$ top cladding. The silicon structures were formed with a 248 nm deep-UV lithography and dry etching. Resistive filaments of titanium nitride with a $3~\upmu\text{m}$ width were deposited on the top cladding with a metal lift off process, used for the on-chip thermal phase shifters. The fabrication of the chip was courtesy of the Cornerstone foundry based at the University of Southampton.
\paragraph{\footnotesize{Phase retrieval}}
In the waveguide nonlinearity measurements, the transmitted power and spectrum of our passively mode-locked Ho-doped fibre pulsed laser (AdValue photonics) were taken \cite{sinclair2019temperature}. Firstly, the pump input pulses are filtered with a single pass monochromator centred at $\lambda = 2.0715~\microns$. The pulses in the time domain are characterised by taking an intensity autocorrelation (Femtochrome FR-103PD) and fitted with a hyperbolic secant-squared ansatz giving an input pulse duration of $\tau = 4.82~\text{ps}$ (see Supplementary Section~S2 for more detail). The laser is then connected to the $17.5-\text{mm}$ spiral with an 9:1 optical tap to monitor the input power. The optical power launched into the chip is controlled with a knife edge variable optical attenuator, allowing the intensity-dependent transmission and spectrum of the output pulse to be measured. For each power launched, a frequency spectrum is recorded on an OSA (Yokogawa AQ6375) at the chip output.

To retrieve nonlinear phase from the output spectrum, sequential Fourier and inverse Fourier transforms are performed to transform between the time and frequency domains (Gerchberg-Saxton algorithm \cite{Fienup:82}). By retaining the phase information, but replacing the amplitude with our measured secant-pulse envelope (time domain) or power spectrum (frequency domain), the algorithm is found to converge on a steady state. Using this algorithm the nonlinear phase corresponding to the measured spectral broadening is retrieved for each pulse. Finally, the phase profiles are fitted with a secant-shaped phase model and combining this with the free carrier dispersion and absorption effects found from fitting the transmission data, the nonlinear phase and hence the waveguide nonlinearity are found.
To determine the value of $n_2$ in Fig.~\ref{fig:motivation}, we calculated the effective modal area using the method from \cite{rukhlenko2012effective}.
In addition to determining the refractive nonlinearity, the transmitted power is also used to determine the TPA parameter by using a fit to the inverse transmission against the input power, as described in Ref. \citenum{PhysRevLett.30.901}.
\paragraph{\footnotesize{Stimulated four-wave mixing}}
We pump the waveguide source with a filtered picosecond pulsed laser centred at $\lambda = 2.0715~\microns$ and a tuneable continuous wave (CW) laser (Sacher Lion) as the stimulating field. The two lasers are combined on a 1:1 fibre beam splitter before the chip with power monitoring optical taps to normalise the stimulated FWM at the chip output. Tuning the frequency of the CW laser, the stimulated FWM from a 7.2-mm spiral is measured on an OSA.
\paragraph{\footnotesize{Detector characterisation}}
Electrical output traces of the RF pulses from the SNSPDs are recorded for several bias settings. The peak voltage of the pulse changes linearly as a function of detector bias and a linear model is fit to this data: 7.46 and 10.84 $\text{mV} \upmu \text{A}^{-1}$ for detectors $A$ and $B$, respectively (Supplementary Fig.~S3). This is combined with a constant offset of 15 and 20 $\text{mV}$, determining the discrimination voltage as a function of bias for detectors $A$ and $B$, respectively. Detectors $A$ and $B$ are connected internally to cryostat ports D4 and D3, respectively.

The transmission of a tuneable CW laser (Sacher Lion) centred at $\lambda = 2.07~\microns$ is measured with a InGaAs photodiode (Thorlabs S148C) at a fibre just before the cryostat to calibrate out any fibre losses between the source and the detectors. A motorised knife edge attenuator is then used to calibrate the knife edge position as a function of transmitted power at the photodiode. Individual transmission of four neutral density filters are then recorded before inserting them in the beam path and reducing the net photon flux down to the single photon regime.

Finally, the photodiode input fibre is connected to the superconducting detector, and singles counts from a Picoquant PicoHarp timetagger are integrated for $0.5~\text{s}$ for each optical power launched. The estimated input and output photon flux is fit with a linear slope model and is used to determine the system detection efficiency (see Supplementary Section~3 for error analysis).

\paragraph{\footnotesize{Correlated photon pair measurement}}

We pump the waveguide source shown in Fig.~\ref{fig:photonpairs}a with a passively mode-locked Ho-doped fibre pulsed laser (AdValue Photonics) centred on $2.0715~\microns$, producing $2.9$-ps pulses, at a $39.4$-MHz repetition rate, with $487~\text{W}$ of peak power coupled into fibre from free space. The laser is filtered with a monochromator in a double-pass configuration (Fig.~\ref{fig:photonpairs}b), producing $5.78$-ps pulses, with $24.4$-W peak power. Filtered pump is then coupled into the waveguide with a V-groove array and vertical grating couplers. After propagating through the waveguide spiral, the signal and idler photons are passively demultiplexed on a fibre beam splitter. Back-to-back single-pass monochromators suppress the pump. Finally, the photon pairs are detected by our SNSPDs. A time-interval analyser (PicoQuant PicoHarp) with a 4-ps resolution correlates detector output pulses. For each power launched, a histogram was integrated for 540 and 180 seconds, in the low and high power regimes, respectively. Each integration was subdivided into 20-second intervals. 

At low pump powers, nonlinear absorption and dispersion play a reduced role, greatly simplifying our model, and allowing us to unambiguously estimate a source efficiency, exclusive of channel losses\cite{rarity1987absolute}. In the low-power regime, SFWM simply scales with $P^2$, where $P$ is the peak pump power. Here, we consider $P<0.5~\mathrm{W}$ to be `low power'. We estimated the on-chip pair-generation efficiency and probability by analysing the pair-generation versus pump-power data of Fig.~\ref{fig:photonpairs}g (net coincidences, $X$) and Supplementary Fig.~S4 (singles of each channel, $C_0$, $C_1$). All three datasets were fit with a polynomial of the form $aP^2 + bP + c$, though with $b=c=0$ for the net coincidences data (as accidentals are already subtracted). The on-chip rate per unit peak-power-squared is then estimated as $a_{C_0} a_{C_1}/a_{X}$, where $a_*$ is the best fit parameter $a$ for the indicated dataset. This efficiency is converted to probability by dividing by the laser repetition rate, and converted to average power by dividing by the square of the effective duty cycle (about $2.6\times10^{-4}$ for our filtered laser).

\paragraph{\footnotesize{Thermo-optic phase modulator calibration}}

We used a 16-bit DC voltage source (Qontrol Systems) to drive the on-chip thermo-optic phase modulators. Injecting CW laser light centred on $\lambda = 2.049~\microns$ (Eblana photonics quantum cascade laser) into the input vertical grating coupler, the optical output of the chip was monitored with a photodiode while the phase-voltage was varied, see Supplementary Fig.~S1. The interferometers voltage-squared versus transmission was fit with a sinusoidal envelope. In the time-reversed HOM experiment, 30 evenly spaced squared-voltages were set on the source interferometer phase shifter.

\end{footnotesize}

\end{document}